\documentclass[a4paper,fleqn,usenatbib,useAMS]{mnras}
\usepackage{epsfig}
\usepackage{graphicx}
\usepackage{amssymb}
\usepackage{amsmath}
\usepackage{rotating}
\usepackage{longtable}

\newcommand{\rv}{{$R(V)$}}
\newcommand{\ebv}{{$E(B-V)$}}
\newcommand{\teff}{{$T_{\rm eff}$}}
\newcommand{\logg}{{$\log g$}}

\newcommand{\fuse}{{\it FUSE}}
\newcommand{\ha}{H$\alpha$}
\newcommand{\mdot}{$\dot{M}$}
\newcommand{\mdotir}{$\dot{M}({\rm IR})$}
\newcommand{\mdotv}{$\dot{M}({\rm Vink})$}

\newcommand{\spit}{{\it Spitzer}}

\newcommand{\kms}{km s$^{-1}$}
\newcommand{\etal}{et al.}
\newcommand{\msunyr}{M$_\odot$~yr$^{-1}$}

\usepackage[T1]{fontenc}
\usepackage{ae,aecompl}

\title[IR Mass Loss Rates for LMC and SMC O Stars]{Mass loss rates from 
mid-IR excesses in LMC and SMC O stars}

\author[D. Massa, A.W. Fullerton, R.K. Prinja]{D. Massa,$^{1}$
\thanks{dmassa@spacescience.org} A.W. Fullerton,$^{2}$ and R.K. Prinja$^{3}$
\\
$^{1}$Space Science Institute, Boulder, CO 80301, USA\\
$^{2}$Space Telescope Science Institute, Baltimore, MD 21218, USA\\
$^{3}$Department of Physics and Astronomy, University College 
London, Gower Street, London WC1E 6BT, UK}

\date{Last updated 2015 May 22; in original form 2013 September 5}

\pubyear{2017}

\begin{document}
\label{firstpage}
\pagerange{\pageref{firstpage}--\pageref{lastpage}}
\maketitle

\begin{abstract}
We use a combination of $BVJHK$ and {\it Spitzer} [3.6], [5.8] and [8.0] 
photometry to determine IR excesses for a sample of 58 LMC and 46 SMC O 
stars. This sample is ideal for determining IR excesses because the very 
small line of sight reddening minimizes uncertainties due to extinction 
corrections.  We use the core-halo model developed by Lamers \& Waters 
(1984a) to translate the excesses into mass loss rates and demonstrate that 
the results of this simple model agree with the more sophisticated CMFGEN 
models to within a factor of 2.  Taken at face value, the derived mass loss 
rates are larger than those predicted by Vink et al.\ (2001), and the 
magnitude of the disagreement increases with decreasing luminosity.  
However, the IR excesses need not imply large mass loss rates.  Instead, 
we argue that they probably indicate that the outer atmospheres of O stars 
contain complex structures and that their winds are launched with much 
smaller velocity gradients than normally assumed.  If this is the case, it 
could affect the theoretical and observational interpretations of the ``weak 
wind'' problem, where classical mass loss indicators suggest that the mass 
loss rates of lower luminosity O stars are far less than expected.  
\end{abstract}


\section{Introduction}\label{sec:intro}

The winds of massive stars power and enrich the ISM, affect the evolution 
of the stars, determine their ultimate fate and the nature of their 
remnants, influence the appearance of the integrated spectra of young, 
massive clusters and star-bursts, and play a major role in the initial 
stages of massive star cluster formation and their subsequent evolution.  
Consequently, reliable measurements of mass loss rates due to stellar winds 
are essential for all of these subjects.  

Stellar winds are driven by radiative pressure on metal lines (Castor \etal\ 
1975, CAK).  However, in recent years, it has become apparent that the winds 
are far more complex than the homogeneous, spherically symmetric flows 
envisioned by CAK.  Instead, they have been shown to contain optically thick 
structures which may be quite small or very large.  Further, these 
structures are thought to have non-monotonic radial velocities.  Co-rotating 
interaction regions (CIRs) (Cranmer \& Owocki, 1996, Lobel \& Blomme, 2008) 
are examples of large structures and wind fragments caused by the line 
deshadowing instability (LDI) (Owocki \etal\ 1988, Sunqvist \etal\ 2011, 
\v{S}urlan \etal\ 2012) are examples of small structures.  Until the 
details of these flows are unraveled, we cannot reliably translate 
observational diagnostics into physical quantities such as mass loss rates.  
To progress, a firm grasp on the underlying physical mechanisms which 
determine the wind structures is required.  The state of affairs can be seen 
in recent literature where the values of observationally derived mass loss 
rates have swung back and forth by factors of 10 or more (Puls \etal\ 2006, 
Massa \etal\ 2003, Fullerton \etal\ 2006, Sunqvist \etal\ 2011, \v{S}urlan 
\etal\ 2012).  

Evidence for large scale wind structure first emerged when the variability 
was observed in \ha\ line profiles (Underhill, 1961, Rosendhal, 1973a, b and 
Ebbets 1982).  This was followed by studies of UV P Cygni line variability 
by several investigators, who examined the behavior of discrete absorption 
components (DACs), which traverse UV wind line profiles and suggest the 
presence of large, coherent structures propagating through the winds (e.g., 
Kaper \etal\ 1999, Prinja \etal\ 2002).  Similar features are observed in 
LMC and SMC O stars (Massa \etal\ 2000) and in planetary nebula central 
stars (Prinja \etal\ 2012), suggesting these structures are a universal 
property of radiatively driven flows. 

Perhaps the most compelling evidence that the winds contain optically thick 
structures was provided by Prinja \& Massa (2010) who used doublet ratios to 
demonstrate that apparently unsaturated wind lines often arise in 
structures that are optically very thick, but cover only a fraction of the 
stellar surface.  Further, Massa \& Prinja (2015) used UV excited state 
wind lines to demonstrate that at least some of these structures are quite 
large and originate very near or on the stellar surface.  Additional 
evidence for large scale structure has been deduced from X-ray variability 
(e.g., Massa \etal\ 2014, Rauw \etal\ 2015).  Models which account for 
optically thick structures have been developed (Sundqvist, Puls \& Feldmeier 
2010, and \v{S}urlan \etal\ 2012), and they provide somewhat better 
descriptions of the observations.  However, one must keep in mind that 
whenever optically thick structures are included in a model, {\em geometry 
matters}.  Therefore, it is essential to constrain the shape of the 
structures as much as possible, and the best way to probe the geometry is 
to examine all of the spectral diagnostics available.  Only when all of 
the available diagnostics have been examined, and a model constructed that 
can simultaneously explain them all, will we be assured that 
observationally determined mass loss rates are meaningful.  Each diagnostic 
provides an important piece of the puzzle. 

The IR fluxes of OB stars present an important diagnostic that has been 
largely neglected.  It was shown early on that emission from OB star winds 
should be detectable at IR and radio wavelengths (Wright \& Barlow 1975, 
Panagia \& Felli 1975).  This realization spawned observations of OB stars 
at near and mid-IR wavelengths (e.g., Castor \& Simon 1983, Abbott \etal\ 
1984), but the results were considered untrustworthy for two reasons.  
First, IR photometric systems were still evolving at the time and poorly 
calibrated.  As a result, only rather large excesses could be trusted.  
Second, accurate reddening corrections are essential for interpreting IR 
excesses, since the excess must be measured relative to the stellar flux at 
a wavelength assumed to be free of wind emission, typically $V$ band 
photometry.  However, there are very few lightly reddened, luminous 
Galactic OB stars, and the exact form of the IR reddening law was poorly 
characterized at the time of the early studies.  Nevertheless, there 
remains strong motivation to study IR excesses since, as Puls \etal\ (2006) 
demonstrated, the wavelength dependence of the mid-IR SED can provide 
important information on the radial dependence of clumping in the wind.

This paper has two major goals. The first is to use near IR (NIR) and 
mid-IR observations of Magellanic cloud O stars to determine their IR 
excesses and compare them to theoretical expectations.  The second is to 
compare the IR mass loss rates of LMC and SMC stars to examine how 
metallicity affects the results.  In \S~\ref{sec:sample} we describe our 
sample of stars.  In \S~\ref{sec:params}, we derive the physical parameters 
of the stars and quantify the influence of interstellar extinction.  In 
\S~\ref{sec:model} we motivate, describe and justify the simplified model 
we use to derive mass loss rates from IR excesses.  In \S~\ref{sec:results} 
we describe how we fit the IR photometry and present our results.  In 
\S~\ref{sec:systematics}, we quantify the sensitivity of the derived mass 
loss rates to various systematic effects.  In \S~\ref{sec:discussion} we 
discuss the implications of our results.  

\section{The Sample and Data}\label{sec:sample}

With the advent of \spit, well calibrated mid-IR observations of the 
Magellanic Clouds became available, thanks to the \spit\ SAGE legacy 
data products provided by Meixner \etal\ (2006) for the LMC and Gordon 
\etal\ (2011) for the SMC.  These data present the opportunity to obtain a 
large, uniform set of IR derived mass loss rates from lightly reddened 
stars, with well-determined luminosities.  Bonanos \etal\ (2009, 2010) took 
advantage of the new data and compiled catalogs by starting with all massive 
stars in the LMC and SMC with high quality spectral classifications, and 
then matching them to entries in the \spit\ and other photometric data 
bases.  The catalogs contain $U$, $B$, $V$, and $I$ from various sources 
and $JHK$\ photometry (primarily from the Two Micron All Sky Survey, 2MASS; 
Skrutskie \etal\ 2006 and the targeted IRSF survey, (Kato \etal\ 2007), 
together with \spit\ IRAC [3.6], [4.5], [5.8] and [8.0] photometry and some 
MIPS [24] photometry (see Bonanos \etal\ for details).  The Bonanos \etal\ 
catalogs contain 341 LMC and 195 SMC O stars with high quality spectral 
types and optical, NIR and \spit\ mid-IR photometry through [4.5].  Bonanos 
\etal\ also demonstrated that the O stars had detectable IR excesses due to 
winds, but did not perform a quantitative analysis of individual stars.  

In this paper, we concentrate on a sub-sample of the Bonanos \etal\ (2009, 
2010) catalogs, namely those O stars which are also in the Blair et al.\ 
(2009) \fuse\ sample.  This will allow direct comparison of results derived 
from different diagnostics in many cases.  We rejected stars later than B0, 
since their winds can contain a significant fraction of neutral hydrogen.  
This fraction can be strongly dependent upon NLTE processes and clumping in 
the wind, both of which introduce unwanted complications into the modeling 
(see Petrov \etal\ 2014).  We also rejected WR stars since their massive 
winds require a full treatment of electron scattering, which we neglect.  
After imposing these restrictions, our sample contained 46 SMC and 58 LMC O 
stars (see Tables~\ref{tab:smcstars} and \ref{tab:lmcstars}).  

We supplemented the CCD based optical photometry listed by Bonanos \etal\ 
with photoelectric $V$ and $B$ photometry from the literature whenever 
possible and assigned errors of 0.03 mag to each.  Priority was given to 
the photoelectric photometry.  We eliminated the \spit\ MIPS data, since 
very few stars were detected at [24].  The $I$ band photometry was also 
eliminated for reasons discussed in \S~\ref{sec:model}, and $U$ band 
photometry was not included since CCD $U$ band photometry (which is all 
that exists for most of the stars) often has calibration issues and it 
was not needed for our purposes.  

The Bonanos \etal\ (2009) calibrations and effective wavelengths were used, 
with two exceptions.  First, the Kato \etal\ (2007) IRSF to 2MASS conversion 
factors were applied to the IRSF photometry.  Second, the $B$ band zero 
magnitude flux was decreased by 4\% with respect to the one listed by 
Bonanos \etal.  This was needed to produce $(B-V)$ indices which agree with 
the Fitzpatrick \& Massa (2005) calibration and to insure that derived 
$E(B-V)$ values are greater than 0.

\section{Stellar Parameters and Reddening}\label{sec:params}

To determine the underlying photospheric flux of each program star and its 
expected theoretical mass loss rate, we must know its physical parameters, 
i.e., mass, effective temperature, luminosity and chemical composition.  
We obtain these from the SMC and LMC spectral type to luminosity, effective 
temperature and mass calibrations provided by Weidner \& Vink (2010). 
Tables~\ref{tab:smcstars} and \ref{tab:lmcstars} summarize the physical 
parameters for the SMC and LMC samples, respectively.  We used the spectral 
types from the Bonanos \etal\ catalogs for most stars, and exceptions are 
noted in the tables.

Observed color excesses, $E(B-V)_{obs}$, were determined using $(B-V)_0$ 
values from TLUSTY model atmospheres (Lanz \& Hubeny 2002) with the 
appropriate \teff, \logg\ and metallicity.  These same models were used to 
determine the photospheric fluxes of the stars.  

To characterize the optical and IR extinction, we adopt the Weingartner \& 
Draine (2001) curves for the SMC and LMC.  As with all other wavelength 
ranges, the form of the IR extinction law is variable (e.g., Fitzpatrick \& 
Massa 2009, Schlafly \etal\ 2016).  However, because reddening is minimal 
in most cases, the exact form of the extinction curve used is not too 
important.  Nevertheless, the effects of variations in $R(V) \equiv 
A(V)/E(B-V)$ are taken into account by allowing $E(B-V)$ to be a free 
parameter in fitting the IR continua (see \S~\ref{sec:results}).  The 
consequences of this action are examined in \S~\ref{sec:systematics}.

\section{The Wind Model}\label{sec:model}
\subsection{Formulation of the model}\label{subsec:model}
In general, the wind density of a smooth spherically symmetric flow from a 
star of radius $R_\star$\ is determined by the wind velocity law and the 
mass loss rate, \mdot.  Typically, the velocity law for a wind with a 
terminal velocity $v_\infty$ is assumed to have the form
\begin{equation}
w = \left(1 - \frac{a}{x}\right)^\beta
\label{eq:vlaw}
\end{equation}
where $w = v/v_\infty$, $x = r/R_\star$, $a = 1 - w_0^{1/\beta}$ and $w_0 
= w(x=1)$.  These wind laws have a maximum velocity gradient at $x = 1$, 
and laws with larger $\beta$ parameters accelerate more slowly.   In the 
following, we adopt $\beta = 1$ and $w_0 = 0.01$, which are typical values 
for OB stars.  Whenever possible, $v_\infty$ values derived from UV 
observations were collected from the literature.  If none were available, 
we derived $v_\infty$ from the stellar parameters and the prescription 
provided by Vink \etal\ (2001).  Their method was also used to calculate 
theoretical mass loss rates, \mdot(Vink).  The mass loss rates derived 
from IR excesses turn out to be more sensitive to the wind parameters than 
the stellar parameters.

The continuity equation relates the wind density and the velocity 
\begin{equation}
\rho = \frac{\dot{M}}{4 \pi R_\star^2 v_\infty x^2 w(x)}  
\label{eq:cont}
\end{equation}
Thus, $\rho$ varies rapidly as $x$ approaches 1; is proportional to \mdot; 
is inversely proportional to $v_\infty$; and, is denser at a given $x$ for 
larger $\beta$.  The IR emission from a wind is dominated by free-free and 
free-bound emission and absorption by H and He.  It can originate very near 
the star (the exact radius depends upon wavelength and wind density).  

Because our goal is to survey several objects in order to determine mean 
properties, identify outliers, contrast differences between the LMC and SMC, 
compare the results with theoretical expectations, and search for trends.  
To accomplish this, we model the observed IR excesses of 104 O stars.  
Consequently, we sought the simplest available model that captures the 
essential physics of IR continuum formation.  A computational fast model 
which suffers only a minimal loss of precision is the one developed by 
Lamers \& Waters (1984a, 1984b, LWa, LWb).  This is a core-halo model 
wherein the flux from a static plane parallel model atmosphere is embedded 
in a stellar wind and both the emission and absorption by the wind material 
are treated in detail.  As is typical, the wind is assumed to be uniform and 
spherically symmetric with a density structure set by the velocity law.  For 
simplicity, it is also assumed that the wind is in LTE at a fixed 
temperature, $T_w$, typically 0.8 -- $0.9 \times T_{eff}$.

We employ the LWa model together with TLUSTY models (with appropriate 
SMC or LMC metallicities) for the underlying photospheres.  In this 
core-halo formulation, the observed flux, $f(\lambda)$, and the flux of the 
underlying photosphere, $f(\lambda)_p$, are related by 
\begin{equation}
f(\lambda) = \left[Z(\lambda)_1 +Z(\lambda)_2 
\frac{f(\lambda)_w}{f(\lambda)_p}\right] f(\lambda)_p \label{eq:landw} 
\end{equation}
where $f(\lambda)_w$ is a Planck function with $T = T_w$, and $Z_1$ and 
$Z_2$ are functions which represent the attenuation of the stellar flux by 
the wind and the emission and self-absorption of the wind, respectively.  
These functions are integrals over the impact parameter, $q$, and can be 
calculated very quickly once the optical depth through the wind as a 
function of impact parameter, $\tau(q)$, is known.  While determining 
$\tau(q)$ can be time consuming, it only has to be done once for a given 
set of wind law parameters, $\beta$ and $w_0$.  Consequently, tables of 
$\tau(q)$ as a function of $q$ can be constructed for each velocity law and 
then scaled by the mass loss rate, terminal velocity and $T_w$.  These can 
be integrated very quickly over $q$ to obtain a specific model.  

When fitting the models to the observations, we use a version of 
equation~(\ref{eq:landw}) which is normalized to the flux at $V$ and 
accounts for extinction, viz, 
\begin{equation}
\frac{f(\lambda)}{f(V)} = \frac{[Z(\lambda)_1 +Z(\lambda)_2 f(\lambda)_w 
    /f(\lambda)_p] f(\lambda)_p 10^{-0.4 A(\lambda)}}{[Z(V)_1 +Z(V)_2 f(V)_w 
    /f(V)_p] f(V)_p10^{-0.4 A(V)}}
\end{equation}
where $f(\lambda)_p$ is a TLUSTY model which depends on $T_{eff}$, 
$\log g$ and metallicity, and $Z_1(\lambda)$ and $Z_2(\lambda)$ depend on 
the velocity law parameters, $\beta$, $w_0$ and $v_\infty$ and \mdot.  
We also assume that $Z(V)_1 = 1$ and $Z(V)_2 = 0$, which gives
\begin{equation}
\frac{f(\lambda)}{f(V)} = \left[Z(\lambda)_1 +Z(\lambda)_2 
      \frac{f(\lambda)_w} {f(\lambda)_p}\right] \frac{f(\lambda)_p}{f(V)_p} 
     10^{-0.4 E(\lambda -V)} 
\end{equation}
In performing the fits, we adopt a logarithmic version of this equation, 
\begin{eqnarray}
  \log \frac{f(\lambda)/f(V)}{f(\lambda)_p/f(V)_p} & = & 
     \log \left[Z(\lambda)_1 +Z(\lambda)_2 \frac{f(\lambda)_w} 
     {f(\lambda)_p}\right]  \nonumber \\ 
& & -0.4 E(B-V) k(\lambda -V)\label{eq:fit}
\end{eqnarray}
where $k(\lambda -V) \equiv E(\lambda -V)/E(B-V)$, so $E(\lambda -V) = 
E(B-V) k(\lambda -V)$.  Note that $k(\lambda -V) < 0$ for $\lambda > 
\lambda_V$.

Following LWa, we do not extend the wind model to wavelengths shorter 
than 1~\micron.  Shorter wavelengths require including the Paschen jump at 
0.82~\micron.  This could introduce sizable uncertainties.  The strength of 
the Paschen jump is much stronger than the Brackett jump (at 1.46~\micron) 
because the populations of the Hydrogen levels increase dramatically with 
decreasing quantum number.  Further, the populations of the lower hydrogen 
levels are strongly affected by NLTE effects, so the exact strength of the 
jump cannot be accurately predicted by the LTE assumption of the LWa model.  
As a result, we bypass the problem by not including $I$ band photometry in 
the fits, which is the only filter strongly affected.  In addition, 
available $I$ band calibrations are not very reliable (see Fitzpatrick \& 
Massa 2005).

Because the LWa core-halo models can be calculated almost instantaneously, 
they are ideal for the current work because we use a non-linear least 
squares fitting routine, where each fit typically involves several tens of 
model calculations, and this has to be done for 104 stars.  In addition, we 
also examine how a variety of constraints and systematic errors affect the 
results, a feat which would be extremely time consuming for more 
sophisticated models. We also ignore the effects of electron scattering, but 
this should be a minor effect for the stars and wavelengths considered here.  

A final simplification used to increase calculation speed was to apply a 
photometric calibration based on effective wavelength, $\lambda_{eff}$.  
We tested the validity of this approach by comparing the fluxes derived 
this way to those determined from integrating over filter response curves.  
Over the small range of intrinsic colors and color excesses in the current 
sample, the effects were all less than 1\%.    

\subsection{Accuracy of the model}\label{subsec:accuracy}
To determine the accuracy of the mass loss rates determined by the simple 
LWa model, we performed the following experiment.  We employed the unclumped 
CMFGEN models calculated by Martins \& Plez (2006) as surrogates for actual 
stars.  These were then fit with LWa models, using eq.~(\ref{eq:fit}).  To 
begin, we constructed a set of Galactic abundance TLUSTY models whose \teff\ 
and $\log g$ values correspond to the CMFGEN models.  For the LWa models, we 
used $\beta = 0.9$ (same as the CMFGEN models used) and assumed $w_0 = 0.01$  
since, unlike the CMFGEN models, the LWa models do not continue into the 
photosphere.  We set $T_w = 0.9 T_{eff}$, since this gave the best overall 
agreement.  The near equality of the wind and stellar temperatures is 
reasonable since, at the wavelengths considered here, the bulk of the wind 
emission comes from very near the stellar surface.  The fits also allowed 
for reddening to be present as well, using a Weingartner \& Draine (2001) 
$R(V) = 3.1$ extinction curve.  This simulates fitting actual data, since 
any difference between the LWa models and the CMFGEN models which has a 
wavelength dependence similar to an extinction curve will be absorbed into 
the measured extinction.  We also allowed for a 2\% error in each point, to 
simulate photometric errors.  

All of the fits were excellent, with reduced $\chi^2 < 1$.  A few aspects 
of the fits are noteworthy.  First, two models (\teff = 32,500K, \mdot 
$\times 10^{6} = 0.011$ and \teff = 37,500K, \mdot $\times 10^{6} = 9.33$) 
do not fit the trends defined by models with similar parameters.  The 
reason for this discord is unknown.  Second, very few of the $E(B-V)$ values 
derived from the fits are larger than 0.01 mag, implying that the model 
distinguishes between reddening and wind excesses very well.  Third, some 
CMFGEN models with very low mass loss rates have IR continua that are {\em 
fainter} than the corresponding TLUSTY model, resulting in small ``negative 
excesses'', which resulted in negative $E(B-V)$ values.  This effect is 
likely related to differences in the structure of the outer atmospheres of 
the CMFGEN and TLUSTY models caused by including the dynamic nature of the 
outer atmosphere in the CMFGEN models.  While interesting, this effect is 
very small, with the magnitude of the negative excesses always less than 
0.01 mag.  

\begin{figure}
\includegraphics[width=1.0\linewidth]{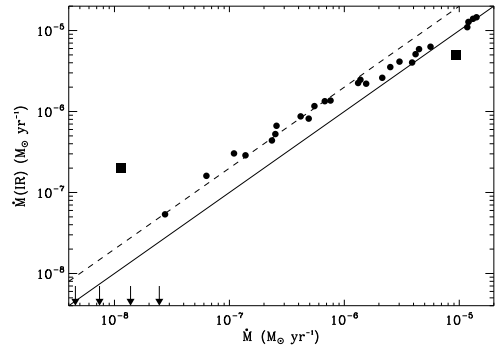} 
\vspace{-2.0in}\caption{Mass loss rates (in \msunyr) determined from LWa 
model fits to the IR continua of the CMFGEN models of Martins \& Plez 
(2006), \mdotir, versus the CMFGEN model mass loss rates, \mdot.  The points 
shown as squares are for CMFGEN models whose fluxes deviate from the trends 
of models with similar physical parameters.  The upper limits are for models 
where the fit could not determine a significant \mdot, when photometric 
errors of 0.02 mag are assigned to each photometry point.  The solid line is 
\mdot$ = $\mdotir, and the dashed line is \mdot$ = $2 \mdotir.  Overall, the 
mass loss rates determined by the simple LWa models recover the CMFGEN rates 
very well for high mass loss rates, but then begin to overestimate the mass 
loss rates for models with smaller \mdot.  Even so, the disagreement is 
typically within a factor of 2 for cases where mass loss rates are 
detectable. 
\label{fig:ratios}}
\end{figure}

Figure~\ref{fig:ratios} summarizes the comparison of the two models.  It 
shows the mass loss rate derived from the LWa model that provides the best 
fit to the IR continuum of the CMFGEN model, \mdot (IR), versus the CMFGEN 
\mdot s.  The \mdot (IR) values are nearly identical to the CMFGEN \mdot s 
for models with large mass loss rates.  For \mdot\ $\lesssim 2 \times 
10^{-6}$~\msunyr, the \mdot (IR)s overestimate the CMFGEN \mdot s by  
about a factor of 2.  Finally, for the smallest \mdot s, the fitting cannot 
detect a significant excess, resulting in \mdot(IR) $\simeq 0$.  

The significant result of this exercise is that if we assume that the 
CMFGEN models provide a good representation of the IR continua of real 
stars, then the simple LWa core halo model faithfully represents actual IR 
continua.  Further, fitting IR continua with the LWa model results in 
\mdot (IR)s that are very accurate for stars with  \mdot\ $\gtrsim 2 \times 
10^{-6}$ \msunyr, but may overestimate the actual \mdot s by a factor of 2 
for \mdot\ $\lesssim 2 \times 10^{-6}$ \msunyr.

\section{Results}\label{sec:results}
To summarize, the following ingredients are used in the fits: Weidner \& 
Vink (2010) tables to translate spectral types into physical parameters; 
TLUSTY models with these parameters to give the photospheric fluxes; and, 
Weingartner \& Draine (2001) extinction curves to characterize the 
extinction.  We examine the implications of these assumptions in 
\S~\ref{sec:systematics}.  
\begin{figure}
\begin{center}
\includegraphics[width=1.0\linewidth]{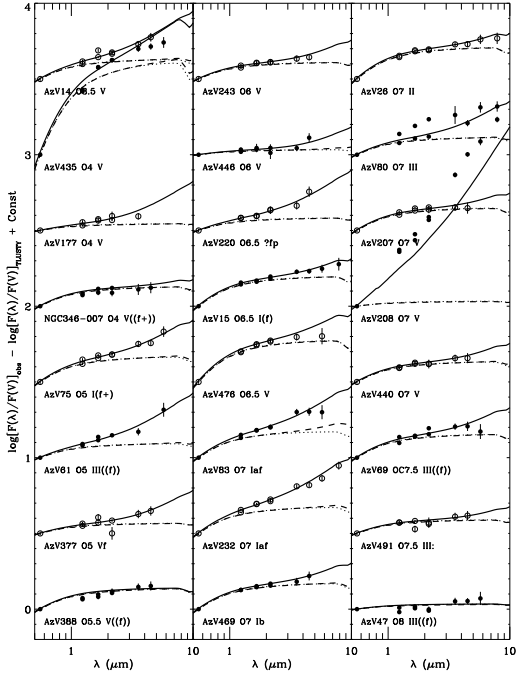} 
\end{center}
\vspace{-0.5in}
\caption{Each plot shows a model fit to the log of the fluxes minus the log 
of the appropriate TLUSTY model, both normalized to $V$.  For each star, the 
observed data are shown as points, with every other curve shown as open or 
filled points to avoid confusion when points of adjoining SEDs overlap.  
The black curve is the model fit, the dotted curve dotted curve shows the 
reddening determined by the fit, and the dashed curve is the reddening plus 
the excess expected from the Vink \etal\ (2001) mass loss rate, \mdotv.  
One $\sigma$ errors (often smaller than the points) are shown and successive 
curves are offset by 0.5 dex from the bottom for display.
\label{fig:smcfits1}}
\end{figure}

\begin{figure}
\begin{center}
\includegraphics[width=1.0\linewidth]{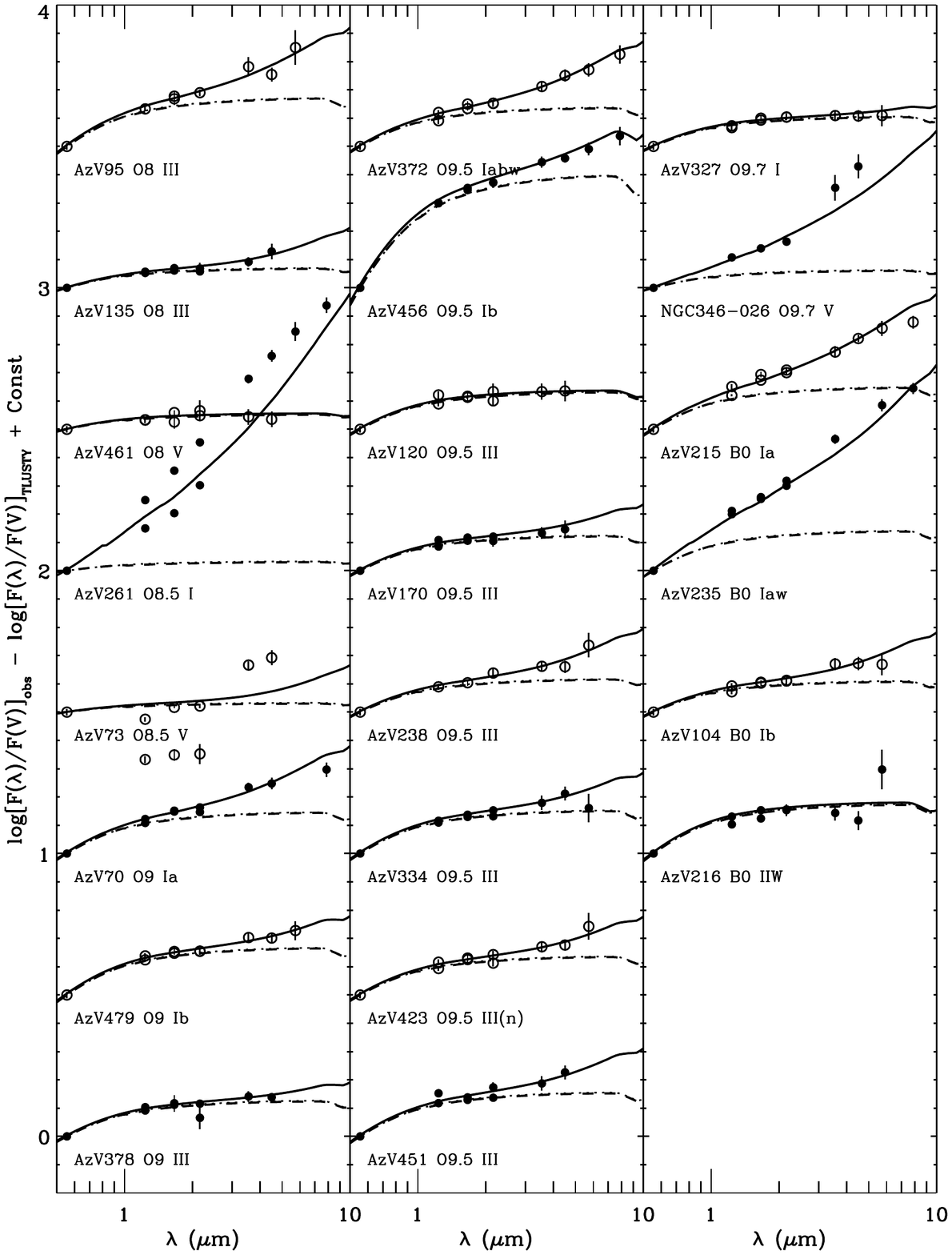}  
\end{center}
\vspace{-0.5in}
\caption{Same as Figure~\ref{fig:smcfits1}.\label{fig:smcfits2}}
\end{figure}

\begin{figure}
\begin{center}
\includegraphics[width=1.0\linewidth]{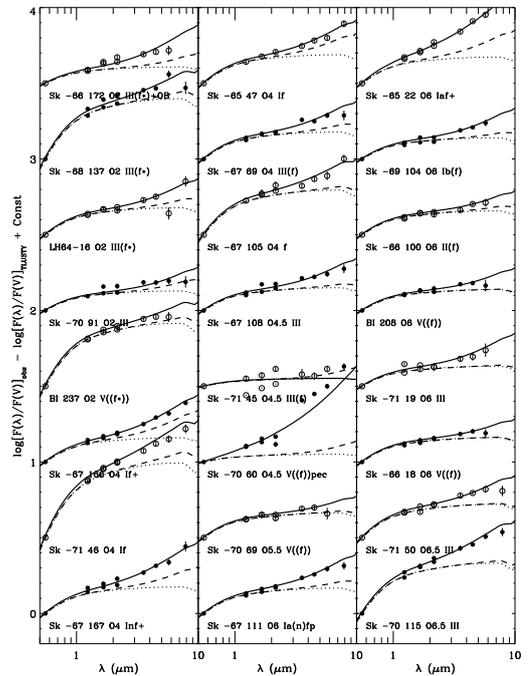} 
\end{center}
\vspace{-0.5in}
\caption{Same as Figure~\ref{fig:smcfits1} for the LMC sample. 
\label{fig:lmcfits1}}
\end{figure}

For each star, the observed and TLUSTY SEDs were normalized by their $V$ 
band fluxes.  The difference, $\log f(\lambda)/f(V) - \log f(\lambda)_p/
f(V)_p$, is the IR excess.  This excess was fit using equation~(
\ref{eq:fit}) and a non-linear least squares routine to determine 2 free 
parameters: \mdot(IR), which is the mass loss rate of the best fitting LWa 
model and $E(B-V)$ (or, equivalently, $R(V)$, see \S~\ref{sec:systematics}).  

\begin{figure}
\begin{center}
\includegraphics[width=1.0\linewidth]{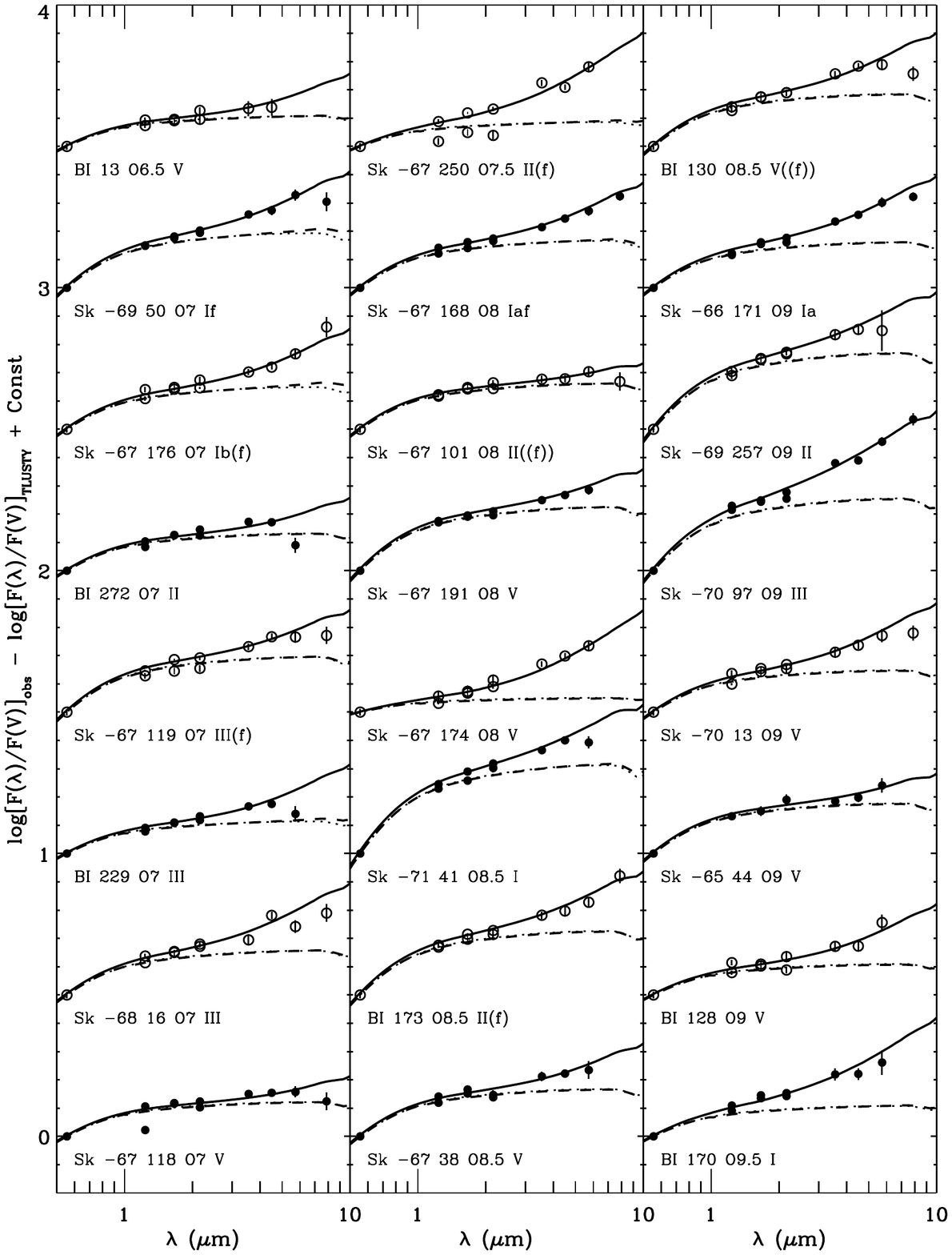}
\end{center}
\vspace{-0.5in}
\caption{Same as Figure~\ref{fig:lmcfits1}. 
\label{fig:lmcfits2}}
\end{figure}

\begin{figure}
\vspace{-1.5in}
\begin{center}
\includegraphics[width=1.0\linewidth]{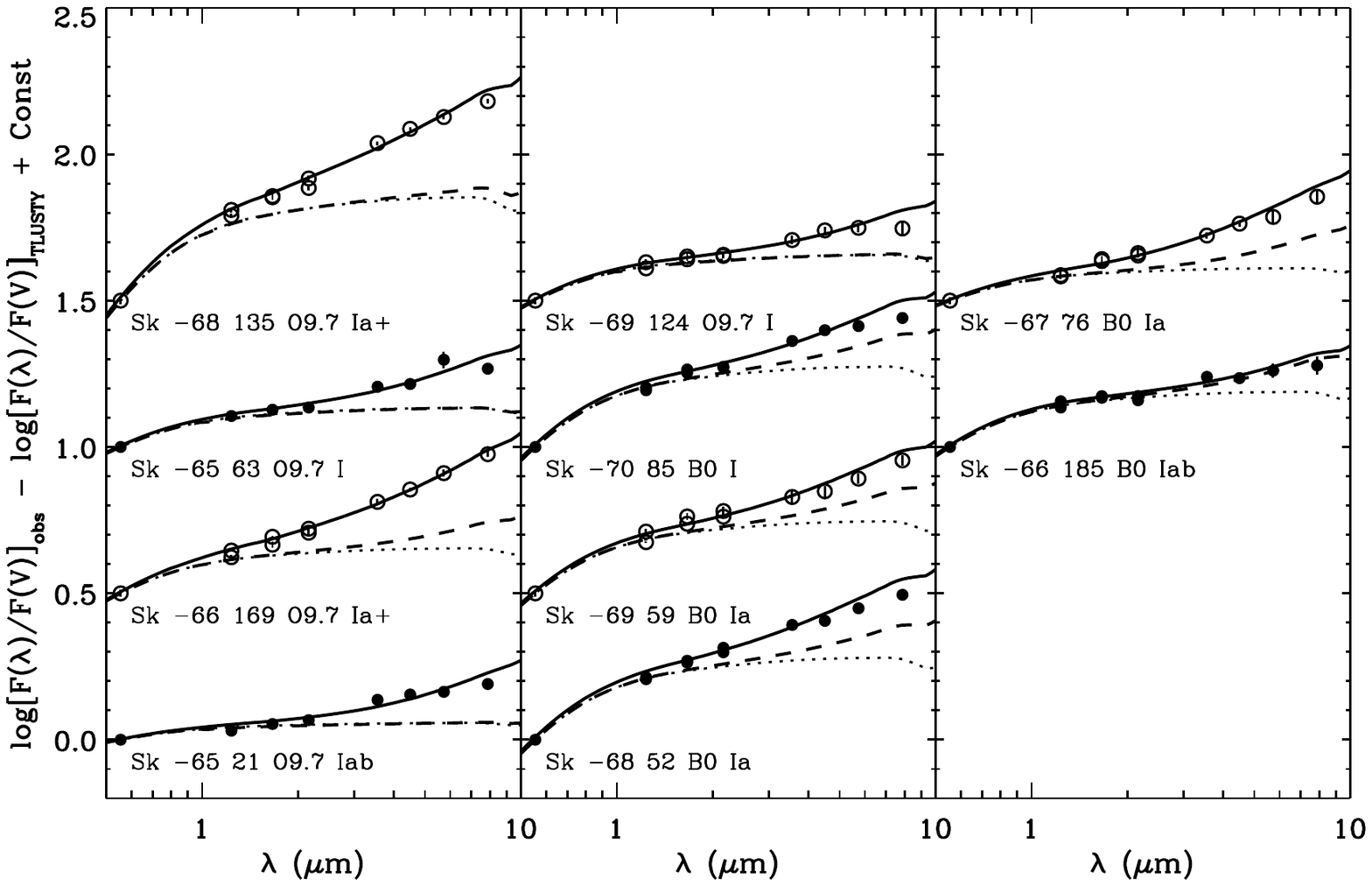} 
\end{center}
\vspace{-0.5in}
\caption{Same as Figure~\ref{fig:lmcfits1}. 
\label{fig:lmcfits3}}
\end{figure}

Figures~\ref{fig:smcfits1} -- \ref{fig:smcfits2} show the SMC fits, 
Figures~\ref{fig:lmcfits1} -- \ref{fig:lmcfits3} show the LMC fits, and 
Tables~\ref{tab:smcwinds} and \ref{tab:lmcwinds} list the results.  Because 
these figures show the differences between the log of the observed fluxes 
and the appropriate TLUSTY model, any shape is due to either reddening or 
wind excess.  The photometry is shown as filled of open points and the best 
fit model for each star is shown as a thick black curve.  The excess that 
would result from the derived reddening and the excess expected from the 
mass loss rate predicted using the Vink \etal\ (2001) formulae, \mdotv,  is 
shown as the dashed curve.  The contribution of reddening to the fits is 
shown as a dotted curve (not always visible because it often coincides with 
the dashed curve).  For each star, the difference between the dotted curve 
and the black curve is the IR excess, and the difference between the dashed 
curve and the black curve is the excess over the IR continuum expected for 
\mdotv.  A few stars, such as AzV~47, show little evidence of either wind 
excess or reddening.  Several, such as AzV~120 and AzV~216, are well fit by 
reddening alone.  For most stars, there is a clear excess relative to pure 
extinction and to the \mdotv\ curves.  

In general, the fits in Figures~\ref{fig:smcfits1} -- \ref{fig:lmcfits3} are 
quite good.  However, four stars: Az~V207, AzV~461, NGC346-026 and AzV~235, 
have extremely large excesses and are poorly fit.  The \mdot(IR) for these 
stars are unreliable since their huge excesses suggest circumstellar disks 
(see Figure~10 in Bonanos \etal\ 2010).  Consequently, stars with 
\mdotir/\mdotv $ > 100$ (more than twice the next largest ratio) are 
identified as probable disks systems in Table~\ref{tab:smcwinds} (all are 
in the SMC).  Notice too, that there is evidence that the IR fluxes varied 
for some stars with large excesses.  For example, the two sets of nearly 
parallel, but off set, points for AzV~243 are from the two $JHK$ surveys, 
which were obtained at different epochs.  

A few other properties of the SEDs are also noteworthy.  For example, 
the SEDs of a few stars, such as AzV435 and Sk$-71^\circ46$, show the 
effects of relatively large reddening.  It is also interesting to contrast 
the excesses and \mdot(IR)s of the O2 star Sk$-70^\circ90$ and the 
O4.5 star Sk$-67^\circ108$.  The former has a smaller excess, but a larger 
\mdot(IR). This is because the O2 star has a much larger $v_\infty$, so the 
wind density (and IR emission) is lower for the same \mdot(IR).  

While the fits for most stars shown in Figures~\ref{fig:smcfits1} -- 
\ref{fig:lmcfits3} appear to have distinct IR excesses, the evidence is 
marginal in some cases (e.g., AzV69, AzV135, BI~208 and Sk$-67^\circ$118).  
This is an important point since even a small IR excess can imply a fairly 
large \mdot(IR), $\sim 10^{-6}$ $M_\odot$~yr$^{-1}$, which can be far 
larger than expected.  To address this problem, all of the stars were fitted 
a second time, with \ebv\ as the only free parameter and \mdot(IR) set to 0.  
We then formed the ratio of the $\chi^2$s for the fits with and without 
mass loss and compared them to an $F$-distribution.  Stars whose ratios 
correspond to a 50\% or more probability of being drawn from the same 
distribution are flagged in Tables~\ref{tab:smcwinds} and \ref{tab:lmcwinds}, 
since there is a good chance that they have no detectable mass loss.  

Tables~\ref{tab:smcwinds} and \ref{tab:lmcwinds} contain several stars whose 
fits result in relatively large $\chi^2$.  Ignoring the probable disks, an 
inspection of Figures~\ref{fig:smcfits1} -- \ref{fig:lmcfits3} reveals that 
nearly all of the fits with $\chi^2 \ge 4$ are due to a large discrepancy 
in the two sets of $JHK$ photometry.  This means that either the photometry 
errors are incorrect, or that the $JHK$ fluxes are variable.  Whatever the 
case, the fits tend to go through the mean of the two sets and usually fit 
the mid-IR quite well.  Since it is the {\it Spitzer} fluxes which determine 
\mdot (IR), we believe that most of these fits are better than their 
$\chi^2$ would indicate.  However, there are two cases that cannot be 
explained as a discordant NIR photometry.  One is AzV~388, whose continuum 
seems to have a distinctly different shape than expected, for reasons that 
are not clear.  The other is AzV~435 whose derived $E(B-V)$ is the largest 
of the entire sample.  Further, it has largest difference between $E(B-V)$ 
and $E(B-V)_{obs}$, suggesting a peculiar extinction curve, whose shape may 
be very different from the one used in the fitting, resulting in a poor fit. 
This case illustrates the difficulties encountered when fitting the IR 
continua of heavily reddened stars.  

\begin{figure}
\begin{center}
\includegraphics[width=1.0\linewidth]{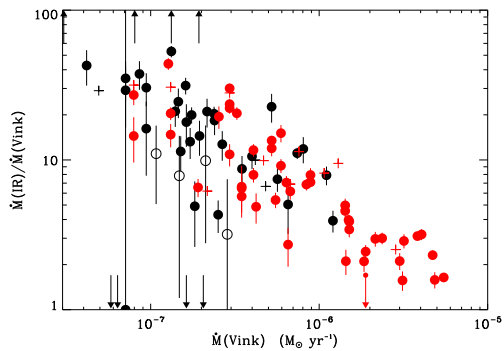}
\end{center}
\vspace{-2.0in}
\caption{Ratios of the mass loss rates determined from IR excesses, 
\mdot(IR), to the theoretical mass loss rates, \mdotv, versus \mdotv.  Red 
and black points are for the LMC and SMC samples, respectively.  Open 
circles represent stars whose measured \mdot(IR)s are consistent with zero, 
downward arrows are for stars whose best fits give \mdot(IR$)=0$, crosses 
are for stars with reduced $\chi^2 > 4$ and upward arrows are for stars 
with very large mass loss rates, probably from disks.
\label{fig:results}}
\end{figure}

Figure~\ref{fig:results} shows the ratio of the derived \mdot(IR) divided by 
\mdotv\ for all of the program stars plotted against \mdotv.  The red 
symbols are for LMC stars and the black symbols for SMC stars.  The downward 
pointing arrows indicate stars whose best fits implied \mdotir~$=0$, stars 
with a 50\% or more probability of having no wind are shown as open symbols, 
and stars whose fits had a $\chi^2 \ge 4$ are shown as crosses.  Two aspects 
of Figure~\ref{fig:results} are worth noting.  The first is that for 
\mdotv~$\gtrsim 10^{-6}\; M_\odot$ yr$^{-1}$, the relation between the 
observed and expected mass loss rates tightens, and \mdotir~$\simeq 
2$~\mdotv.  The second is that for \mdotv~$\lesssim 10^{-7}\; M_\odot$ 
yr$^{-1}$, even when the non-detections, probable non-detections and stars 
with obvious disks are ignored, more than half of the stars still have solid 
detections, implying that the large measured excesses are quite real. 

\section{Systematic effects}\label{sec:systematics}
This section examines how the derived mass loss rates are affected by errors 
in: spectral classifications; the neglect of wind emission on the $B$ and 
$V$ photometry; variations in the extinction curves; the assumed wind 
temperature; and, the velocity law parameters.  

Changing the spectral and luminosity classes (and, hence, the stellar 
parameters) changes $f(\lambda)_p$ and \mdotv.  Nevertheless, simulations 
showed that the derived \mdotir\ were not very sensitive to classification 
errors of $\pm 1$ spectral or luminosity class and changed by less than 20\% 
in both cases. 

Comparisons to CMFGEN models (\S~\ref{sec:model}), suggest that our 
assumption that $B$ and $V$ are unaffected by wind emission is reasonable 
in most cases.  However, in the few instances where the wind emission is 
strong enough to affect the optical photometry, it will contaminate $V$ 
more than $B$.  This creates an intrinsic $E(B-V)$, causing an over 
correction for extinction which, in turn, leaves less excess to be 
accounted for by \mdot(IR), resulting in an underestimate of \mdot(IR).  
Since our major concern is effects that might lead us to over estimate 
\mdot(IR), this issue is not considered further. 

\begin{figure}
\begin{center}
\includegraphics[width=1.0\linewidth]{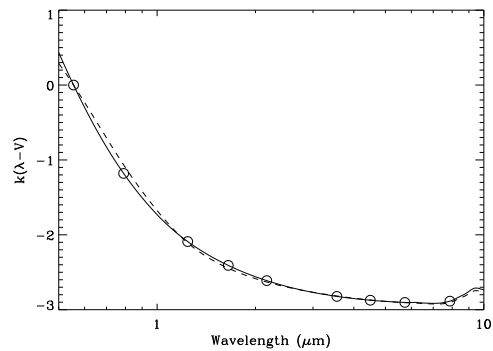}
\vspace{-2.1in}
\caption{Comparison of Weingartner \& Draine (2001) extinction curves for 
$R(V) = 3.1$ (black curve) and $R(V) = 5.5$ (dashed curve).  The $R(V) = 
5.5$ has been divided by 1.7 to demonstrate how very different curves can 
appear proportional for wavelengths longer than $V$.\label{fig:extinction}}
\end{center}
\end{figure}

Variations in the extinction law are another concern.  
Figure~\ref{fig:extinction} demonstrates how changing $E(B-V)$ or $R(V)$ are 
equivalent over the wavelength range of interest.  It shows two Weingartner 
\& Draine (2001) Galactic curves: one for $R(V) = 3.1$ and one for $R(V) = 
5.5$ divided by 1.7.  The locations of the photometric bands are also 
indicated.  The figure shows that for wavelengths longer than $V$, rescaling 
an extinction curve with one value of $R(V)$ results in a good approximation 
of an extinction curve with a different $R(V)$.  However, for large values 
of $E(B-V)$, the differences can become important, introducing errors of 0.1 
mag or more.  That is why it is best to derive IR excesses for stars with 
small color excesses. 

\begin{figure}
\begin{center}
\includegraphics[width=1.0\linewidth]{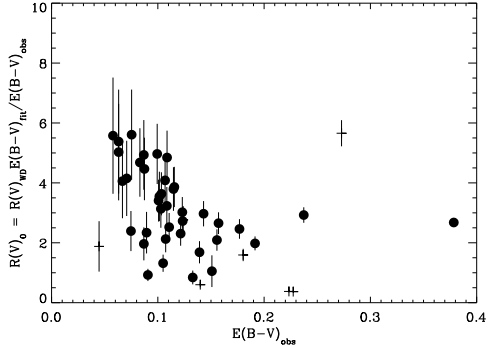}
\end{center}
\vspace{-2.0in}
\caption{
Values of \rv\ derived from the fits plotted against \ebv$_{obs}$ for the 
SMC sample.  Solid points have a reduced $\chi^2 < 4$ and crosses have 
$\chi^2 \geq 4$.  The plot suggests a foreground contribution to the 
extinction of \ebv$\simeq 0.05$\ mag from dust with a rather large \rv, 
which then melds into SMC dust with a much smaller \rv.  Both components 
may be rather patchy, accounting for the large scatter.\label{fig:rv_smc}}
\end{figure}

The degeneracy of \ebv\ and \rv\ was the motivation for allowing \ebv\ to be 
a free parameter when fitting the SEDs instead of simply setting it equal to 
$E(B-V)_{obs}$.  This accommodates possible variations in \rv.  If the 
actual value of \rv\ along the line of sight, \rv$_0$, differs from the 
assumed Weingartner \& Draine (2001) value, \rv$_{WD}$, then it should be 
possible to recover \rv$_0$, from the relation
\begin{equation}
E(B-V)_{obs} R(V)_{WD} = E(B-V)_{fit} R(V)_0 
\end{equation}
where $E(B-V)_{fit}$ is the excess obtained from the fit.   
Figure~\ref{fig:rv_smc} shows \rv$_0$ plotted against \ebv$_{obs}$ for the 
SMC sample.  Considering that all of the excesses are small and that the 
errors are large, we see that there is a general trend for \rv$_0$ to 
decrease with increasing $E(B-V)_{obs}$.  This is consistent with the sight 
lines passing through a small amount of foreground, Galactic dust, with an 
\rv $\sim 4$ and then passing through more and more SMC dust with $R(V) 
\sim 2.5$.  Although the line of sight to the SMC was not included in 
Schlafly \etal\ (2017), they do detect $R(V) \sim 4$ for high latitude 
Galactic dust in nearby fields.  Further, $R(V) = 2.5$ is consistent with 
the SMC $R(V)$ determined by Gordon \& Clayton (1998).  Keeping in mind 
that neither the foreground nor the SMC dust are probably perfectly uniform, 
the general trend appears to verify our assumption that the value of \rv\ 
is changing from one line of sight to the next, and the amount of change 
depends on the relative amounts of Galactic and SMC dust encountered.  

Assigning an incorrect temperature to the wind can affect the results.  The 
wind temperature could be much higher than 0.9\teff\ if shocks and their 
associated X-rays heat the wind to temperatures $\sim 10^6$K as suggested 
by Cassinelli \etal\ (2001).  However, increasing $T_{w}$ to $10^6$~K only 
reduces \mdot(IR) by 30\%.  

Finally, we examined how changing the parameters in the velocity law, 
eq.~(\ref{eq:vlaw}), affects the derived mass loss rates.  Changing $w_0$ 
over the range $0.005 \leq w_0 \leq 0.02$ (the most commonly used values) 
changed the derived \mdotir s by 20\% or less, which is small compared to 
the observed disagreement between theory and observation.  

It is particularly important to examine the effect of changing $v_\infty$ 
for the less luminous stars.  For most of these stars, observed values are 
not available so we rely on the values predicted by Vink \etal\ (2001).  
When observed values are available, they are often much less than the 
predicted ones, but this could be a systematic effect.  It is difficult to 
measure $v_\infty$ in less luminous stars, since their wind lines are 
typically asymmetric, lacking a distinctive blue edge.  This makes the full 
extent of the wind absorption hard to measure and easy to underestimate.  
However, experiments show that reducing $v_\infty$ by a factor of 2 reduces 
\mdotir\ by 40\%.  While substantial, this is too little to explain the 
observed \mdotir / \mdotv\ ratios.  

\begin{figure}
\begin{center}
\includegraphics[width=1.0\linewidth]{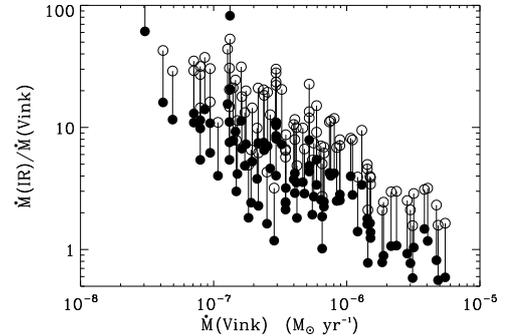}
\end{center}
\vspace{-2.0in}
\caption{
The effect of changing $\beta$ on the derived mass loss rates for the sample
shown in Figure~\ref{fig:results}.  The plot shows \mdotir/\mdotv\ as a 
function of \mdotv.  Open points are \mdotir\ derived using $\beta =1$, 
filled points are \mdotir\ derived using $\beta = 2.5$, and lines connect 
points for the same star.  
\label{fig:beta}}
\end{figure}

In contrast to the other parameters, $\beta$\ can have a major effect on 
the derived \mdot(IR)s (as noted by LWa).  Changing $\beta$ from 1 to 2.5 
reduces the inferred \mdot\ by a factor of 2.7.  Figure~\ref{fig:beta} 
demonstrates this effect.  It arises because a larger $\beta$ reduces the 
velocity gradient (increasing the density) near the surface of the star, 
where most of the emission occurs.  We return to this issue in the next 
section.  

\section{Discussion}\label{sec:discussion}

Our results can be summarized as follows: mass loss rates determined from IR 
excesses are larger than expected, increasing from about a factor of 2 for 
the most luminous stars to about 40 for the least luminous stars with 
detectable IR excesses; and, the LMC and the SMC results are similar.  
Because we have shown that \mdotir\ derived by the LWa models can 
overestimate the actual \mdot\ by a factor of 2 for stars with $\dot{M} 
\lesssim 2 \times 10^{-6}$ \msunyr, these results are consistent with 
\mdotv~$\lesssim$ \mdotir $\lesssim$ 2\mdotv\ for \mdotv~$\gtrsim 2 
\times 10^{-6}$ \msunyr, growing to 5 \mdotv~$\lesssim$ \mdotir~$\lesssim 
20$~\mdotv\ for the least luminous stars.  Thus, if we interpret the IR 
excesses in terms of mass loss, they infer rates that are equal to or 
much larger than theoretical predictions.  This is in stark contrast to 
recent mass loss rates determined from UV wind lines, \ha\ and X-ray 
diagnostics, which suggest rates smaller than the Vink \etal\ (2001) 
predictions by a factor of 2 to 3 (see Martins \& Palacios 2017 for a 
summary).

However, it is well known that increasing the mass loss rate is not the only 
way to increase the IR emission.  In this section, we argue that the large 
IR excesses and their implied large \mdot(IR)s are not the result of large 
mass loss rates.  Instead, we attribute them to density and velocity 
structures near the base of the wind.  We also consider the implications of 
the scatter in Figure~\ref{fig:results} and the fact that points from stars 
in both galaxies are interspersed.  

Throughout the discussion, we emphasize that the IR excesses in our sample 
originate very near the stellar surface.  To see this, consider the 
effective radii of the IR excesses.  Using a $\beta = 1$ velocity law and 
our longest wavelength, 8.0$\mu$m, even the strongest winds in our sample 
(aside from the obvious disks) result in a value of the LWa parameter 
$E_\nu$ (their eq.~6) that is less than 0.01.  This implies an effective 
radius which is only a few percent larger than the stellar radius (see LWa 
Figure~4).  

At O star temperatures, the intensity of the IR wind emission is 
proportional to the density squared, and there are two ways to enhance 
this for a fixed mass loss rate.  One is to collect most of the wind mass 
into structures or clumps with enhanced densities.  The other is to reduce 
the velocity gradient, which then increases the density through the 
continuity equation.  Since, as outlined in \S~\ref{sec:intro}, evidence 
for structure in O star winds abounds, including indications that this 
structure originates near the stellar surface, we suspect that this 
structure accounts for much of the large excesses.  The notion that we are 
seeing effects that originate near the stellar surface is reinforced by the 
fact that the largest discrepancies occur for stars with lower expected mass 
loss rates, where we expect to see nearly to the stellar surface.  

It is possible to make our results a bit more quantitative.  First, consider 
the case where all of the excess is due to density inhomogeneities.  Abbott 
\etal\ (1981) developed a simple model for how clumping enhances the 
observed emission.  It assumes the wind consists of two components whose 
density ratio is $x$, where $0 \leq x \leq 1$, and the fraction of the wind 
volume occupied by the higher density is $f_V$.  The geometry of the 
structures is unspecified.  Setting $x = 0$ gives the largest enhancements 
for fixed $f_V$, with the ratio of the observed to actual mass loss rates 
being simply $f_V^{-1/2}$.  In our case, this ratio is \mdotir/\mdotv.  For 
stars with smaller \mdotv, this ratio varies between 5 and 20 (where an 
overestimate of 2 by \mdotir\ is assumed).  Thus, if the entire discrepancy 
is attributed to clumping, stars with weak winds must have all of the wind 
material confined to between 0.3 and 4\% of the wind volume.  This is in 
contrast to stars with more massive the winds, where Massa \& Prinja (2015) 
showed that the wind structures near the photosphere are quite large and to 
recent \ha\ results, described below.  

Next, consider the effects of the velocity gradient.  Figure~\ref{fig:beta} 
shows that increasing $\beta$ from 1 to 2.5 (which corresponds to 
reducing the velocity gradient at $x=1$ by a factor of 7.5 for $w_0 = 
0.01$), decreases the inferred \mdotir\ by a factor of 2.7.  Thus, a large 
reduction in the velocity gradient can also strongly influence the mass 
loss inferred from the IR flux.  

It is interesting to compare our results with the recent \ha\ mass loss 
rates determined by Ram\'{i}rez-Agudelo \etal\ (2017).  For stars more 
luminous than $\log L/L_\odot \gtrsim 5.6$, (which corresponds roughly to 
\mdotv $\gtrsim 10^{-6}$ \msunyr) both approaches require similar volume 
filling factors to bring the observational mass loss rates into accord with 
the theoretical ones.  However, at lower luminosity, the IR mass loss rates 
are much larger than those determined from \ha.  This can arise if the 
velocity gradient is small, so that the opacity at continuum wavelengths is 
much less than in a line like \ha.  In this case, the IR emission probes 
more deeply into the wind and, as we saw above, a slow acceleration can 
greatly enhance the IR emission.  Therefore, to avoid extremely large 
values of $f_V$, it seems that some combination of strong density clumping 
and a small acceleration at the base of the wind are required to produce 
the observed IR excesses.  Given that the IR emission in stars expected to 
have low mass loss rates originates in the poorly understood transition 
region between the photosphere and wind, our results should not be too 
surprising.  Instead, we should view the IR as providing important 
constraints on the structure of the photosphere -- wind interface.

A few other aspects of Figure~\ref{fig:results} are also of interest. First, 
the large intrinsic scatter in \mdotir\ for stars with smaller \mdotv\ 
suggests that either the IR excesses are affected by physical parameters 
beyond those used to determine \mdotv\ (e.g., rotation, magnetic fields, 
interacting binary winds or incipient disks), or that the excesses in these 
stars are variable.  Second, points from both the SMC and LMC overlap, 
implying that the physical origin of the process causing the additional 
excesses is independent of metallicity, and consistent with the Vink et al.\ 
(2001) treatment of different metallicities.  

Our results could also have bearing on the ``weak wind problem'' (Martins 
\etal\ 2005).  If the excesses for low luminosity stars are due to extremely 
compact structures and small velocity gradients near the stellar surfaces, 
then many of the diagnostics used to interpret weak winds could be strongly 
affected.  

\section*{Acknowledgements}
DM acknowledges support from ADP Grant No.\ NNX14AB30G to SSI.  This 
publication makes use of data products from the 2MASS, which is a joint 
project of the University of Massachusetts and the Infrared Processing and 
Analysis Center/California Institute of Technology, funded by the National 
Aeronautics and Space Administration and the National Science Foundation.


\newpage

\begin{onecolumn}
\begin{center}
\begin{longtable}{lllrrrrrr}
\caption{SMC Stellar Properties}\label{tab:smcstars}\\
\hline
Name        & Sp Ty  & ref & $V$ & $(B-V)$ & \teff & $\log L/L_\odot$ & $M/M_\odot$ & $R/R_\odot$\\ 
\hline
AzV14           & O3-4 V               & B10 &   13.77 &   -0.19 &   44338 &   5.44 &   44.35 &   8.91 \\
AzV435          & O4 V                 & B10 &   14.13 &   -0.07 &   43292 &   5.40 &   41.15 &   8.88 \\
AzV177          & O4 V                 & B10 &   14.53 &   -0.21 &   43292 &   5.40 &   41.15 &   8.88 \\
NGC346-007      & O4 V((f+))           & B10 &   14.02 &   -0.24 &   43292 &   5.40 &   41.15 &   8.88 \\
AzV75           & O5 I(f+)             & B10 &   12.79 &   -0.16 &   38715 &   5.81 &   54.79 &  17.76 \\
AzV61           & O5 III((f))          & B10 &   13.54 &   -0.18 &   38881 &   5.68 &   42.91 &  15.20 \\
AzV377          & O5 Vf                & B10 &   14.59 &   -0.25 &   41200 &   5.31 &   35.38 &   8.86 \\
AzV388          & O5.5 V((f))          & B10 &   14.12 &   -0.21 &   40154 &   5.27 &   32.78 &   8.87 \\
AzV243          & O6 V                 & B10 &   13.87 &   -0.22 &   39108 &   5.22 &   30.33 &   8.89 \\
AzV446          & O6 V                 & B10 &   14.59 &   -0.24 &   39108 &   5.22 &   30.33 &   8.89 \\
AzV220          & O6.5 ?fp             & B10 &   14.50 &   -0.22 &   35929 &   5.67 &   46.83 &  17.64 \\
AzV15           & O6.5 I(f)            & B10 &   13.17 &   -0.21 &   35929 &   5.67 &   46.83 &  17.64 \\
AzV476          & O6.5 V               & B10 &   13.52 &   -0.09 &   38062 &   5.18 &   28.01 &   8.92 \\
AzV83           & O7 Iaf               & B10 &   13.58 &   -0.13 &   35001 &   5.62 &   44.90 &  17.64 \\
AzV232          & O7 Iaf               & B10 &   12.36 &   -0.20 &   35001 &   5.62 &   44.90 &  17.64 \\
AzV469          & O7 Ib                & B10 &   13.20 &   -0.22 &   35001 &   5.62 &   44.90 &  17.64 \\
AzV26           & O7 II                & B10 &   12.55 &   -0.20 &   35226 &   5.54 &   39.36 &  15.86 \\
AzV80           & O7 III               & M95 &   13.32 &   -0.13 &   35451 &   5.46 &   33.83 &  14.27 \\
AzV207          & O7 V                 & B10 &   14.37 &   -0.22 &   37016 &   5.13 &   25.80 &   8.97 \\
AzV208          & O7 V                 & B10 &   14.15 &   -0.10 &   37016 &   5.13 &   25.80 &   8.97 \\
AzV440          & O7 V                 & B10 &   14.64 &   -0.20 &   37016 &   5.13 &   25.80 &   8.97 \\
AzV69           & OC7.5 III((f))       & B10 &   13.35 &   -0.22 &   34593 &   5.41 &   32.05 &  14.08 \\
AzV491          & O7.5 III:            & M95 &   14.72 &   -0.20 &   34593 &   5.41 &   32.05 &  14.08 \\
AzV47           & O8 III((f))          & B10 &   13.38 &   -0.26 &   33736 &   5.35 &   30.39 &  13.92 \\
AzV95           & O8 III               & B10 &   13.83 &   -0.19 &   33736 &   5.35 &   30.39 &  13.92 \\
AzV135          & O8 III               & B10 &   13.96 &   -0.23 &   33736 &   5.35 &   30.39 &  13.92 \\
AzV461          & O8 V                 & B10 &   14.61 &   -0.21 &   34924 &   5.05 &   21.63 &   9.10 \\
AzV261          & O8.5 I               & P09 &   13.88 &   -0.07 &   32215 &   5.49 &   40.52 &  17.81 \\
AzV73           & O8.5 V               & M95 &   14.08 &   -0.17 &   33878 &   5.00 &   19.63 &   9.20 \\
AzV70           & O9 Ia                & P09 &   12.38 &   -0.17 &   31287 &   5.44 &   39.35 &  17.93 \\
AzV479          & O9 Ib                & B10 &   12.48 &   -0.15 &   31287 &   5.44 &   39.35 &  17.93 \\
AzV378          & O9 III               & B10 &   13.88 &   -0.24 &   32021 &   5.25 &   27.29 &  13.65 \\
AzV451          & O9 V                 & M02 &   14.15 &   -0.23 &   32832 &   4.96 &   17.64 &   9.31 \\
AzV372          & O9.5 Iabw            & P09 &   12.63 &   -0.18 &   30358 &   5.40 &   38.22 &  18.07 \\
AzV456          & O9.5 Ib              & B10 &   12.89 &    0.09 &   30358 &   5.40 &   38.22 &  18.07 \\
AzV238          & O9.5 II              & P09 &   13.77 &   -0.22 &   30761 &   5.30 &   31.99 &  15.64 \\
AzV423          & O9.5 II(n)           & P09 &   13.28 &   -0.19 &   30761 &   5.30 &   31.99 &  15.64 \\
AzV120          & O9.5 III             & B10 &   14.56 &   -0.23 &   31163 &   5.19 &   25.76 &  13.55 \\
AzV170          & O9.5 III             & B10 &   14.09 &   -0.23 &   31163 &   5.19 &   25.76 &  13.55 \\
AzV334          & O9.5 III             & B10 &   13.81 &   -0.21 &   31163 &   5.19 &   25.76 &  13.55 \\
AzV327          & O9.7 I               & B10 &   13.25 &   -0.22 &   29987 &   5.38 &   37.78 &  18.14 \\
AzV215          & B0 Ia                & P09 &   12.69 &   -0.09 &   29430 &   5.35 &   37.10 &  18.25 \\
AzV235          & B0 Iaw               & P09 &   12.20 &   -0.18 &   29430 &   5.35 &   37.10 &  18.25 \\
AzV104          & B0 Ib                & P09 &   13.17 &   -0.16 &   29430 &   5.35 &   37.10 &  18.25 \\
AzV216          & B0 IIW               & M02 &   14.32 &   -0.17 &   29868 &   5.25 &   30.65 &  15.67 \\
NGC346-026      & B0 IV                & E09 &   14.87 &   -0.14 &   30740 &   4.87 &   13.65 &   9.59 \\
\hline
\multicolumn{9}{l}{{\bf Notes:} B10: Bonanos et al. 2010; P09: Penny \& Gies (2009); M95: Massey \etal\ (1995);}\\ 
\multicolumn{9}{l}{M02: Massey \etal\ (2002)}\\
\end{longtable}
\end{center}
\clearpage

\begin{center}
\begin{longtable}{lllrrrrrr}
\caption{LMC Stellar Properties}\label{tab:lmcstars}\\ 
\hline
Name        & Sp Ty  & ref   & $V$ & $(B-V)$ & \teff & $\log L/L_\odot$ & $M/M_\odot$ & $R/R_\odot$\\ 
\hline
Sk -66 172      & O2 III(f*)+OB        & B09 &   13.13 &   -0.12 &   48849 &   5.93 &   84.09 &  12.92 \\
Sk -68 137      & O2 III(f*)           & B09 &   13.35 &   -0.08 &   48849 &   5.93 &   84.09 &  12.92 \\
LH64-16         & ON2 III(f*)          & B09 &   13.67 &   -0.22 &   48849 &   5.93 &   84.09 &  12.92 \\
Sk -70 91       & O2 III               & B09 &   12.78 &   -0.23 &   48849 &   5.93 &   84.09 &  12.92 \\
BI 237          & O2 V((f*))           & B09 &   13.98 &   -0.12 &   51269 &   5.82 &   79.66 &  10.24 \\
Sk -67 166      & O4 If+               & B09 &   12.27 &   -0.22 &   41809 &   5.92 &   60.68 &  17.39 \\
Sk -71 46       & O4 If                & B09 &   13.25 &   -0.09 &   41809 &   5.92 &   60.68 &  17.39 \\
Sk -67 167      & O4 Inf+              & B09 &   12.54 &   -0.19 &   41809 &   5.92 &   60.68 &  17.39 \\
Sk -65 47       & O4 If                & B09 &   12.51 &   -0.18 &   41809 &   5.92 &   60.68 &  17.39 \\
Sk -67 69       & O4 III(f)            & B09 &   13.09 &   -0.16 &   43985 &   5.70 &   54.76 &  12.13 \\
Sk -67 105      & O4 f                 & B09 &   12.42 &   -0.15 &   43985 &   5.70 &   54.76 &  12.13 \\
Sk -67 108      & O4-5 III             & B09 &   12.56 &   -0.20 &   43985 &   5.70 &   54.76 &  12.13 \\
Sk -71 45       & O4-5 III(f)          & B09 &   11.47 &   -0.11 &   43985 &   5.70 &   54.76 &  12.13 \\
Sk -70 60       & O4-5 V((f))pec       & B09 &   13.85 &   -0.19 &   46395 &   5.56 &   52.67 &   9.30 \\
Sk -70 69       & O5.5 V((f))          & B09 &   13.94 &   -0.27 &   43958 &   5.43 &   43.40 &   8.93 \\
Sk -67 111      & O6 Ia(n)fp           & B09 &   12.57 &   -0.20 &   37415 &   5.75 &   46.07 &  17.78 \\
Sk -65 22       & O6 Iaf+              & B09 &   12.07 &   -0.19 &   37415 &   5.75 &   46.07 &  17.78 \\
Sk -69 104      & O6 Ib(f)             & B09 &   12.10 &   -0.21 &   37415 &   5.75 &   46.07 &  17.78 \\
Sk -66 100      & O6 II(f)             & B09 &   13.26 &   -0.21 &   38268 &   5.60 &   42.32 &  14.39 \\
BI 208          & O6 V((f))            & B09 &   13.96 &   -0.24 &   41521 &   5.30 &   36.13 &   8.63 \\
Sk -71 19       & O6 III               & B09 &   14.27 &   -0.20 &   39121 &   5.46 &   38.57 &  11.67 \\
Sk -66 18       & O6 V((f))            & B09 &   13.50 &   -0.20 &   41521 &   5.30 &   36.13 &   8.63 \\
Sk -71 50       & O6.5 III             & B09 &   13.45 &   -0.30 &   39121 &   5.46 &   38.57 &  11.67 \\
Sk -70 115      & O6.5 III             & B09 &   12.24 &   -0.10 &   39121 &   5.46 &   38.57 &  11.67 \\
BI 13           & O6.5 V               & B09 &   13.85 &   -0.19 &   41521 &   5.30 &   36.13 &   8.63 \\
Sk -69 50       & O7 If                & B09 &   13.26 &   -0.13 &   35218 &   5.66 &   41.41 &  18.15 \\
Sk -67 176      & O7 Ib(f)             & B09 &   11.82 &   -0.16 &   35218 &   5.66 &   41.41 &  18.15 \\
BI 272          & O7 II                & B09 &   13.28 &   -0.22 &   35953 &   5.50 &   37.41 &  14.49 \\
Sk -67 119      & O7 III(f)            & B09 &   13.33 &   -0.21 &   36689 &   5.34 &   33.41 &  11.58 \\
BI 229          & O7 III               & B09 &   12.95 &   -0.17 &   36689 &   5.34 &   33.41 &  11.58 \\
Sk -68 16       & O7 III               & B09 &   12.96 &   -0.15 &   36689 &   5.34 &   33.41 &  11.58 \\
Sk -67 118      & O7 V                 & B09 &   12.99 &   -0.20 &   39084 &   5.17 &   30.24 &   8.40 \\
Sk -67 250      & O7.5 II(f)           & B09 &   12.68 &   -0.17 &   35953 &   5.50 &   37.41 &  14.49 \\
Sk -67 168      & O8 Iaf               & B09 &   12.08 &   -0.17 &   33021 &   5.57 &   37.71 &  18.68 \\
Sk -67 101      & O8 II((f))           & B09 &   12.63 &   -0.17 &   33639 &   5.40 &   33.44 &  14.70 \\
Sk -67 191      & O8 V                 & B09 &   13.46 &   -0.21 &   36647 &   5.04 &   25.12 &   8.24 \\
Sk -67 174      & O8 V                 & B09 &   11.67 &   -0.18 &   36647 &   5.04 &   25.12 &   8.24 \\
Sk -71 41       & O8.5 I               & B09 &   12.84 &   -0.07 &   33021 &   5.57 &   37.71 &  18.68 \\
BI 173          & O8.5 II(f)           & B09 &   13.00 &   -0.14 &   33639 &   5.40 &   33.44 &  14.70 \\
Sk -67 38       & O8.5 V               & P09 &   13.72 &   -0.23 &   36647 &   5.04 &   25.12 &   8.24 \\
BI 130          & O8.5 V((f))          & B09 &   12.55 &   -0.22 &   36647 &   5.04 &   25.12 &   8.24 \\
Sk -66 171      & O9 Ia                & B09 &   12.19 &   -0.15 &   30824 &   5.49 &   34.39 &  19.39 \\
Sk -69 257      & O9 II                & B09 &   12.49 &   -0.08 &   31324 &   5.29 &   29.71 &  15.07 \\
Sk -70 97       & O9 III               & B09 &   13.33 &   -0.23 &   31825 &   5.10 &   25.03 &  11.71 \\
Sk -70 13       & O9 V                 & B09 &   12.35 &   -0.15 &   34210 &   4.91 &   20.15 &   8.15 \\
Sk -65 44       & O9 V                 & B09 &   13.65 &   -0.21 &   34210 &   4.91 &   20.15 &   8.15 \\
BI 128          & O9 V                 & B09 &   13.82 &   -0.25 &   34210 &   4.91 &   20.15 &   8.15 \\
BI 170          & O9.5 I               & B09 &   13.09 &   -0.17 &   30824 &   5.49 &   34.39 &  19.39 \\
Sk -68 135      & ON9.7 Ia+            & B09 &   11.36 &    0.00 &   30824 &   5.49 &   34.39 &  19.39 \\
Sk -65 63       & O9.7 I               & B09 &   12.56 &   -0.16 &   30824 &   5.49 &   34.39 &  19.39 \\
Sk -66 169      & O9.7 Ia+             & B09 &   11.56 &   -0.13 &   30824 &   5.49 &   34.39 &  19.39 \\
Sk -65 21       & O9.7 Iab             & B09 &   12.02 &   -0.16 &   30824 &   5.49 &   34.39 &  19.39 \\
Sk -69 124      & O9.7 I               & B09 &   12.81 &   -0.18 &   30824 &   5.49 &   34.39 &  19.39 \\
Sk -70 85       & B0 I                 & B09 &   12.30 &   -0.10 &   28627 &   5.40 &   30.85 &  20.34 \\
Sk -69 59       & B0 Ia                & P09 &   12.13 &   -0.12 &   28627 &   5.40 &   30.85 &  20.34 \\
Sk -68 52       & B0 Ia                & B09 &   11.54 &   -0.07 &   28627 &   5.40 &   30.85 &  20.34 \\
Sk -67 76       & B0 Ia                & P09 &   12.42 &   -0.13 &   28627 &   5.40 &   30.85 &  20.34 \\
Sk -66 185      & B0 Iab               & B09 &   13.11 &   -0.19 &   28627 &   5.40 &   30.85 &  20.34 \\
\hline
\multicolumn{9}{l}{{\bf Notes:} B09: Bonanos e\etal\ (2009); P09: Penny \& 
Gies (2009)}\\
\end{longtable}
\end{center}

\begin{center}
\begin{longtable}{lccccccrl}
\caption{SMC Wind Properties} \label{tab:smcwinds} \\ \hline
Name         & $v_\infty$ & ref & \mdot(Vink) & \mdot(IR) & $E(B-V)$ & $E(B-V)_{obs}$ & $\chi^2$ & \\
 & \kms &  & \multicolumn{2}{c}{$10^{-6} M_\odot$ yr$^{-1}$} & \multicolumn{2}{c}{mag} & & \\
\hline
AzV14           & 2000 & M04 &  0.40 &  4.22 $\pm$  0.75 &  0.12 $\pm$  0.013 &  0.16 &  1.84 & \\
AzV435          & 1500 & M07 &  0.48 &  3.20 $\pm$  0.50 &  0.55 $\pm$  0.013 &  0.27 &  6.09 & \\
AzV177          & 2650 & M05 &  0.24 &  4.39 $\pm$  0.88 &  0.04 $\pm$  0.009 &  0.13 &  3.49 & \\
NGC346-007      & 2300 & M07 &  0.28 &  0.91 $\pm$  1.21 &  0.12 $\pm$  0.006 &  0.10 &  1.53 &  * \\
AzV75           & 2100 & M07 &  1.10 &  8.73 $\pm$  1.25 &  0.15 $\pm$  0.008 &  0.16 &  0.96 & \\
AzV61           & 2025 & M09 &  0.80 &  9.53 $\pm$  1.89 &  0.08 $\pm$  0.014 &  0.14 &  0.96 & \\
AzV377          & 2350 & M04 &  0.19 &  2.82 $\pm$  0.74 &  0.06 $\pm$  0.010 &  0.09 &  0.87 & \\
AzV388          & 1935 & M07 &  0.21 &  0.00 $\pm$ 47.27 &  0.12 $\pm$  0.005 &  0.12 & 10.04 & \\
AzV243          & 2125 & M07 &  0.15 &  1.72 $\pm$  0.45 &  0.10 $\pm$  0.005 &  0.11 &  1.18 & \\
AzV446          & 1400 & M05 &  0.25 &  1.08 $\pm$  0.27 &  0.03 $\pm$  0.000 &  0.09 &  0.44 & \\
AzV220          & 2267 & VVV &  0.52 & 11.83 $\pm$  2.58 &  0.07 $\pm$  0.015 &  0.09 &  0.35 & \\
AzV15           & 2125 & M07 &  0.57 &  4.21 $\pm$  0.74 &  0.18 $\pm$  0.004 &  0.10 &  0.41 & \\
AzV476          & 2646 & VVV &  0.09 &  2.86 $\pm$  0.71 &  0.25 $\pm$  0.007 &  0.24 &  1.28 & \\
AzV83           &  940 & M07 &  1.21 &  4.74 $\pm$  0.78 &  0.16 $\pm$  0.011 &  0.18 &  0.90 & \\
AzV232          & 1400 & M09 &  0.74 &  8.22 $\pm$  0.71 &  0.16 $\pm$  0.007 &  0.11 &  2.39 & \\
AzV469          & 1550 & M07 &  0.65 &  3.30 $\pm$  1.20 &  0.15 $\pm$  0.008 &  0.09 &  0.15 & \\
AzV26           & 2150 & M07 &  0.35 &  3.04 $\pm$  0.66 &  0.19 $\pm$  0.004 &  0.11 &  1.70 & \\
AzV80           & 1550 & E04 &  0.42 &  4.17 $\pm$  0.57 &  0.10 $\pm$  0.007 &  0.18 & 12.63 & \\
AzV207          & 2000 & M05 &  0.11 &  1.19 $\pm$  0.64 &  0.13 $\pm$  0.005 &  0.10 &  0.90 &  * \\
AzV208          & 2537 & VVV &  0.08 & 57.90 $\pm$  0.48 &  0.03 $\pm$  0.000 &  0.22 &  9.72 &  D \\
AzV440          & 1300 & M05 &  0.18 &  0.90 $\pm$  0.42 &  0.12 $\pm$  0.008 &  0.12 &  0.40 & \\
AzV69           & 1800 & M07 &  0.27 &  3.38 $\pm$  0.76 &  0.14 $\pm$  0.007 &  0.09 &  2.40 & \\
AzV491          & 2167 & VVV &  0.21 &  2.09 $\pm$  1.50 &  0.08 $\pm$  0.008 &  0.11 &  0.82 &  * \\
AzV47           & 2140 & VVV &  0.16 &  0.00 $\pm$  0.00 &  0.03 $\pm$  0.000 &  0.04 & 18.12 & \\
AzV95           & 1700 & M07 &  0.22 &  4.55 $\pm$  0.99 &  0.16 $\pm$  0.012 &  0.11 &  0.52 & \\
AzV135          & 2140 & VVV &  0.16 &  2.91 $\pm$  0.92 &  0.06 $\pm$  0.005 &  0.07 &  0.45 & \\
AzV461          & 2310 & VVV &  0.06 &  0.00 $\pm$  0.00 &  0.05 $\pm$  0.005 &  0.10 &  0.61 & \\
AzV261          & 2179 & VVV &  0.19 & 72.30 $\pm$  0.74 &  0.03 $\pm$  0.000 &  0.23 & 40.01 &  D \\
AzV73           & 2190 & VVV &  0.05 &  1.43 $\pm$  0.28 &  0.03 $\pm$  0.000 &  0.14 & 58.06 & \\
AzV70           & 1450 & M07 &  0.24 &  4.84 $\pm$  0.55 &  0.13 $\pm$  0.006 &  0.12 &  1.99 & \\
AzV479          & 2159 & VVV &  0.15 &  3.57 $\pm$  0.80 &  0.15 $\pm$  0.004 &  0.14 &  0.53 & \\
AzV378          & 2078 & VVV &  0.09 &  1.52 $\pm$  0.78 &  0.12 $\pm$  0.004 &  0.06 &  0.40 & \\
AzV451          & 2063 & VVV &  0.04 &  1.78 $\pm$  0.47 &  0.15 $\pm$  0.006 &  0.08 &  1.90 & \\
AzV372          & 1550 & M07 &  0.16 &  5.06 $\pm$  0.64 &  0.13 $\pm$  0.008 &  0.11 &  0.65 & \\
AzV456          & 1450 & M07 &  0.17 &  3.49 $\pm$  0.43 &  0.36 $\pm$  0.004 &  0.38 &  0.16 & \\
AzV238          & 1200 & P96 &  0.17 &  2.27 $\pm$  0.53 &  0.10 $\pm$  0.011 &  0.07 &  0.41 & \\
AzV423          & 2111 & VVV &  0.09 &  3.22 $\pm$  0.70 &  0.12 $\pm$  0.006 &  0.10 &  0.90 & \\
AzV120          & 2039 & VVV &  0.07 &  0.00 $\pm$  0.00 &  0.12 $\pm$  0.007 &  0.06 &  0.74 & \\
AzV170          & 2039 & VVV &  0.07 &  2.07 $\pm$  0.90 &  0.11 $\pm$  0.005 &  0.06 &  1.26 & \\
AzV334          & 2039 & VVV &  0.07 &  2.49 $\pm$  0.74 &  0.14 $\pm$  0.006 &  0.08 &  0.44 & \\
AzV327          & 1500 & E04 &  0.15 &  1.16 $\pm$  0.98 &  0.10 $\pm$  0.005 &  0.07 &  0.47 &  * \\
AzV215          & 1400 & M07 &  0.13 &  7.03 $\pm$  0.61 &  0.14 $\pm$  0.007 &  0.19 &  3.31 & \\
AzV235          & 1400 & M07 &  0.13 & 16.98 $\pm$  1.62 &  0.13 $\pm$  0.016 &  0.10 &  1.50 &  D \\
AzV104          & 1340 & M07 &  0.14 &  2.96 $\pm$  0.63 &  0.10 $\pm$  0.006 &  0.12 &  0.69 & \\
AzV216          & 2077 & VVV &  0.06 &  0.00 $\pm$  0.00 &  0.16 $\pm$  0.003 &  0.12 &  1.83 & \\
NGC346-026      & 1780 & VVV &  0.03 &  5.92 $\pm$  1.42 &  0.06 $\pm$  0.028 &  0.15 &  2.52 &  D\\
\hline
\multicolumn{9}{l}{* -- Consistent with no wind; D -- Probable disk; $v_\infty$ in brackets are 
theoretical values.}\\ 
\multicolumn{9}{l}{{\bf Notes:} E04: Evans \etal\ 2004; M04: Massey \etal\ (2004); M05: Massey \etal\ (2005); M09: } \\  
\multicolumn{9}{l}{Massey \etal\ (2009); M07: Mokiem \etal\ (2007); P96: Puls \etal\ (1996); VVV: Vink 
\etal\ (2001)}\\
\end{longtable}
\end{center}

\clearpage
\begin{center}
\begin{longtable}{lccccccrl}
\caption{LMC Wind Properties} \label{tab:lmcwinds} \\ \hline
Name         & $v_\infty$ & ref & \mdot(Vink) & \mdot(IR) & $E(B-V)$ & $E(B-V)_{obs}$ & $\chi^2$ & \\
 & \kms &  & \multicolumn{2}{c}{$10^{-6} M_\odot$ yr$^{-1}$} & \multicolumn{2}{c}{mag} & & \\ \hline
Sk -66 172      & 3100 & C16 &  3.19 &  9.16 $\pm$ 0.99 &  0.11 $\pm$ 0.007 &  0.23 &  3.33 & \\
Sk -68 137      & 3400 & C16 &  2.84 &  7.17 $\pm$ 0.83 &  0.39 $\pm$ 0.005 &  0.27 &  6.58 & \\
LH64-16         & 3250 & C16 &  3.01 &  6.35 $\pm$ 0.82 &  0.18 $\pm$ 0.005 &  0.13 &  2.05 & \\
Sk -70 91       & 3150 & C16 &  3.12 &  4.90 $\pm$ 0.73 &  0.13 $\pm$ 0.005 &  0.12 &  2.48 & \\
BI 237          & 3400 & C16 &  1.85 &  3.88 $\pm$ 0.58 &  0.41 $\pm$ 0.005 &  0.24 &  1.12 & \\
Sk -67 166      & 1900 & C16 &  5.49 &  9.01 $\pm$ 0.61 &  0.16 $\pm$ 0.006 &  0.11 &  0.56 & \\
Sk -71 46       & 2431 & VVV &  4.05 & 12.89 $\pm$ 1.07 &  0.49 $\pm$ 0.006 &  0.24 &  2.33 & \\
Sk -67 167      & 2150 & C16 &  4.71 & 10.91 $\pm$ 0.79 &  0.17 $\pm$ 0.006 &  0.14 &  1.99 & \\
Sk -65 47       & 2100 & C16 &  4.85 &  7.67 $\pm$ 0.95 &  0.19 $\pm$ 0.012 &  0.15 &  0.36 & \\
Sk -67 69       & 2500 & C16 &  1.89 &  4.61 $\pm$ 0.49 &  0.18 $\pm$ 0.005 &  0.18 &  1.20 & \\
Sk -67 105      & 3001 & VVV &  1.51 &  5.85 $\pm$ 0.62 &  0.29 $\pm$ 0.005 &  0.19 &  0.74 & \\
Sk -67 108      & 3001 & VVV &  1.51 &  5.18 $\pm$ 0.59 &  0.15 $\pm$ 0.006 &  0.14 &  2.71 & \\
Sk -71 45       & 2500 & C16 &  1.89 &  0.00 $\pm$-0.00 &  0.05 $\pm$ 0.000 &  0.23 & 86.30 & \\
Sk -70 60       & 2300 & C16 &  1.30 & 12.30 $\pm$ 0.21 &  0.05 $\pm$ 0.000 &  0.16 & 18.28 & \\
Sk -70 69       & 2750 & C16 &  0.65 &  1.78 $\pm$ 0.51 &  0.16 $\pm$ 0.005 &  0.08 &  0.74 & \\
Sk -67 111      & 2000 & C16 &  2.36 &  7.09 $\pm$ 0.63 &  0.17 $\pm$ 0.005 &  0.11 &  1.16 & \\
Sk -65 22       & 1350 & C16 &  3.83 & 11.88 $\pm$ 0.68 &  0.17 $\pm$ 0.008 &  0.12 &  0.70 & \\
Sk -69 104      & 2159 & VVV &  2.15 &  6.39 $\pm$ 0.73 &  0.14 $\pm$ 0.004 &  0.10 &  2.88 & \\
Sk -66 100      & 2075 & C16 &  1.44 &  3.04 $\pm$ 0.58 &  0.16 $\pm$ 0.004 &  0.11 &  0.73 & \\
BI 208          & 3050 & VVV &  0.35 &  2.24 $\pm$ 0.41 &  0.14 $\pm$ 0.004 &  0.10 &  0.85 & \\
Sk -71 19       & 2633 & VVV &  0.67 &  4.65 $\pm$ 0.80 &  0.13 $\pm$ 0.006 &  0.13 &  4.68 & \\
Sk -66 18       & 3050 & VVV &  0.35 &  1.98 $\pm$ 0.44 &  0.16 $\pm$ 0.005 &  0.14 &  0.56 & \\
Sk -71 50       & 2633 & VVV &  0.67 &  4.16 $\pm$ 0.50 &  0.22 $\pm$ 0.006 &  0.03 &  1.25 & \\
Sk -70 115      & 2200 & C16 &  0.84 &  5.76 $\pm$ 0.50 &  0.33 $\pm$ 0.007 &  0.23 &  1.92 & \\
BI 13           & 3050 & VVV &  0.35 &  2.30 $\pm$ 0.87 &  0.11 $\pm$ 0.007 &  0.15 &  0.72 & \\
Sk -69 50       & 2064 & VVV &  1.43 &  7.09 $\pm$ 0.76 &  0.19 $\pm$ 0.005 &  0.18 &  0.78 & \\
Sk -67 176      & 2064 & VVV &  1.43 &  6.52 $\pm$ 0.58 &  0.15 $\pm$ 0.005 &  0.15 &  1.73 & \\
BI 272          & 3400 & C16 &  0.47 &  4.63 $\pm$ 0.89 &  0.13 $\pm$ 0.004 &  0.09 &  4.84 & \\
Sk -67 119      & 2494 & VVV &  0.41 &  3.24 $\pm$ 0.36 &  0.19 $\pm$ 0.004 &  0.11 &  3.98 & \\
BI 229          & 1950 & C16 &  0.55 &  2.98 $\pm$ 0.35 &  0.11 $\pm$ 0.005 &  0.15 &  2.88 & \\
Sk -68 16       & 2494 & VVV &  0.41 &  4.74 $\pm$ 0.45 &  0.15 $\pm$ 0.005 &  0.17 &  2.86 & \\
Sk -67 118      & 2854 & VVV &  0.22 &  1.34 $\pm$ 0.28 &  0.12 $\pm$ 0.006 &  0.13 &  4.87 & \\
Sk -67 250      & 2291 & VVV &  0.76 &  8.62 $\pm$ 0.75 &  0.09 $\pm$ 0.007 &  0.14 & 12.16 & \\
Sk -67 168      & 1979 & VVV &  0.89 &  6.30 $\pm$ 0.47 &  0.16 $\pm$ 0.004 &  0.13 &  1.55 & \\
Sk -67 101      & 2300 & C16 &  0.42 &  2.06 $\pm$ 0.47 &  0.16 $\pm$ 0.003 &  0.14 &  1.08 & \\
Sk -67 191      & 1950 & C16 &  0.19 &  1.25 $\pm$ 0.17 &  0.22 $\pm$ 0.004 &  0.12 &  0.20 & \\
Sk -67 174      & 2645 & VVV &  0.13 &  4.02 $\pm$ 0.15 &  0.05 $\pm$ 0.000 &  0.15 &  4.22 & \\
Sk -71 41       & 1979 & VVV &  0.89 &  7.05 $\pm$ 0.78 &  0.31 $\pm$ 0.005 &  0.23 &  2.92 & \\
BI 173          & 2850 & C16 &  0.32 &  6.65 $\pm$ 0.64 &  0.22 $\pm$ 0.005 &  0.17 &  1.57 & \\
Sk -67 38       & 2645 & VVV &  0.13 &  1.94 $\pm$ 0.35 &  0.17 $\pm$ 0.007 &  0.10 &  1.25 & \\
BI 130          & 2645 & VVV &  0.13 &  2.69 $\pm$ 0.26 &  0.18 $\pm$ 0.007 &  0.11 &  2.83 & \\
Sk -66 171      & 1885 & VVV &  0.52 &  7.04 $\pm$ 0.52 &  0.16 $\pm$ 0.005 &  0.14 &  1.51 & \\
Sk -69 257      & 2059 & VVV &  0.25 &  4.94 $\pm$ 0.84 &  0.27 $\pm$ 0.008 &  0.22 &  0.58 & \\
Sk -70 97       & 2193 & VVV &  0.13 &  5.58 $\pm$ 0.50 &  0.25 $\pm$ 0.009 &  0.07 &  1.31 & \\
Sk -70 13       & 2392 & VVV &  0.08 &  2.51 $\pm$ 0.24 &  0.15 $\pm$ 0.005 &  0.17 &  4.61 & \\
Sk -65 44       & 2392 & VVV &  0.08 &  1.15 $\pm$ 0.38 &  0.18 $\pm$ 0.012 &  0.11 &  0.36 & \\
BI 128          & 2392 & VVV &  0.08 &  2.15 $\pm$ 0.31 &  0.11 $\pm$ 0.007 &  0.07 &  2.02 & \\
BI 170          & 1700 & C16 &  0.59 &  8.95 $\pm$ 1.18 &  0.11 $\pm$ 0.009 &  0.12 &  2.00 & \\
Sk -68 135      & 1050 & C16 &  1.07 &  8.75 $\pm$ 0.45 &  0.35 $\pm$ 0.009 &  0.29 &  4.61 & \\
Sk -65 63       & 1885 & VVV &  0.52 &  6.25 $\pm$ 0.73 &  0.13 $\pm$ 0.010 &  0.13 &  1.26 & \\
Sk -66 169      &  800 & C16 &  1.49 &  5.94 $\pm$ 0.30 &  0.15 $\pm$ 0.006 &  0.16 &  2.08 & \\
Sk -65 21       & 1700 & C16 &  0.59 &  5.43 $\pm$ 0.58 &  0.06 $\pm$ 0.010 &  0.13 &  2.52 & \\
Sk -69 124      & 1600 & C16 &  0.64 &  4.51 $\pm$ 0.48 &  0.15 $\pm$ 0.005 &  0.11 &  1.08 & \\
Sk -70 85       & 1765 & VVV &  0.29 &  6.52 $\pm$ 0.53 &  0.22 $\pm$ 0.006 &  0.18 &  1.94 & \\
Sk -69 59       & 1765 & VVV &  0.29 &  6.93 $\pm$ 0.60 &  0.19 $\pm$ 0.005 &  0.16 &  3.96 & \\
Sk -68 52       & 1765 & VVV &  0.29 &  8.24 $\pm$ 0.52 &  0.22 $\pm$ 0.005 &  0.21 &  4.12 & \\
Sk -67 76       & 1765 & VVV &  0.29 &  8.84 $\pm$ 0.64 &  0.06 $\pm$ 0.006 &  0.15 &  3.39 & \\
Sk -66 185      & 1765 & VVV &  0.29 &  3.20 $\pm$ 0.54 &  0.13 $\pm$ 0.004 &  0.09 &  1.58 & \\
\hline 
\multicolumn{9}{l}{{\bf Notes:} C16: Crowther \etal\ (2016); VVV: Vink 
\etal\ (2001)}\\
\end{longtable}
\end{center}
\end{onecolumn}
\end{document}